\documentclass{aa}
\usepackage{epsf}

\newcommand{\kms}  {  {\rm km  \ s}^{-1}	}
\newcommand{\Mpc} {  {\rm Mpc} 			}

\newcommand{\hMpc} {  h^{-1}{\rm Mpc}  		}
\newcommand{\M}    {  {{\cal M}	        }	}
\newcommand{\Mvir} {  {{\cal M}_{vir}   }	}
\newcommand{\tcr}  {  {t_{cr}	        }	}
\newcommand{\mlim} {  {m_{\rm lim}	}	}
\newcommand{\Mlim} {  {M_{\rm lim}      }	}

\newcommand{\ltsima} {$\; \buildrel < \over \sim \;$}
\newcommand{\gtsima} {$\; \buildrel > \over \sim \;$}
\newcommand{\simlt}  {\lower.5ex\hbox{\ltsima}}
\newcommand{\simgt}  {\lower.5ex\hbox{\gtsima}}
\begin{document}

\thesaurus{11(04.03.1; 04.19.1; 11.03.1; 11.12.2; 12.12.1}

\title{Loose groups of galaxies in the Perseus--Pisces survey}
\author{	R. Trasarti--Battistoni }
\institute{	Dipartimento di Fisica dell'Universit\`a,  
		Via Celoria 16, I-20133 Milano, Italy}
\thanks{\emph{Present address:}	
		Ludwig--Maximilians--Universit\"at, Sektion Physik,
		Theresienstr. 95, D-80333 M\"unchen, Germany}
\offprints{Roberto Trasarti--Battistoni, roberto@stat.physik.uni-muenchen.de}

\date{Received date / Accepted date}
\titlerunning{Loose Groups in PPS}
\authorrunning{R.~Trasarti--Battistoni}
\maketitle

\begin{abstract}

We present a large catalog
\footnote{
The group catalog presented here is available 
via anonymous ftp to cdsarc.u--strasbg.fr (130.79.128.5) or 
via http://cdsweb.u-strasbg.fr/Abstract.html.
Upon request, the author can provide other group catalogs in electronic form.
} 
of loose groups of galaxies in the Southern Galactic Hemisphere, 
selected from the Perseus--Pisces redshift Survey (PPS). 
Particular care is taken in order to obtain group samples 
as homogeneous as possible to previously published catalogs. 
All our catalogs contain about 200 groups, 
significantly more than in most previous studies 
where group samples were obtained from galaxy data sets 
of comparable quality to (but smaller extent than) PPS. 
Groups are identified with the adaptive Friends--Of--Friends (FOF) algorithm of
Huchra \& Geller (1982), with suitable normalizations 
$D_0=0.231 \ h^{-1}{\rm Mpc}$ and $V_0=350 \ {\rm km \ s}^{-1}$ 
at $cz_0=1000 \ {\rm km \ s}^{-1}$.
The luminosity function (LF) normalization 
$\phi_*=0.02 \ h^3 \ \Mpc^{-3}$ appropriate for PPS yields 
a number density threshold $\delta n/n \approx 180$ for the adopted $D_0$, 
instead of $\delta n/n \approx 80$ used in previous studies of other samples. 
However, the customary choice of $D_0$ obtained (through the LF) from 
a fixed mass overdensity $\delta \rho / \rho=80$, well motivated in theory, 
suffers from 
important observational uncertainties 
and sample--to--sample variations of the LF normalization,
and from
major uncertainties in the relation between 
galaxy density $n$ and mass density $\rho$.
We discuss how to self--consistently match 
FOF parameters among different galaxy samples. 
We then separately vary several FOF and sample parameters,
and discuss their effect on group properties.
Loose groups in PPS nicely trace the large scale structure (LSS)
in the parent galaxy sample.
The group properties vary little with different redshift corrections, 
redshift cut--off, and galaxy LF, but
are rather sensitive to the adopted links $D_0$ and $V_0$.
More precisely, the typical group size (velocity dispersion) is 
linearly related to the adopted distance (velocity) link, while
it is rather insensitive to the adopted velocity (distance) link.
Physical properties of groups in PPS 
and in directly comparable samples show good agreement.
There is a complex interplay among LSS features, sample depth,
FOF grouping procedure, and group properties.

\end{abstract}

\section{Introduction}
\label{sez:introduction}

Galaxies and clusters probe the Large Scale Structure (LSS) 
of the matter distribution in the Universe at various scales of
mass, spatial separation, and density.
The comparison of such different regimes allows to gain a significant insight 
into the relation among invisible and luminous matter.
Galaxy groups can be regarded as systems intermediate between galaxies and
galaxy clusters. 
They provide constraints through two different routes:
as galaxy systems, through their internal properties
(velocity dispersion $\sigma_{v}$, radius $R$, etc.);
as LSS tracers, through their ``external'' properties
(abundance, clustering, fraction of grouped galaxies, etc.).
Groups therefore yield quite useful counterchecks 
to models based on galaxy and cluster data.

The main target of this work is to
present a large and homogeneous catalog of galaxy groups
in the Southern Galactic Hemisphere,
extracted from the Perseus--Pisces redshift Survey 
(PPS hereafter; Giovanelli \& Haynes 1993, 
Wegner et al. 1993, and references therein).
The total number of groups is $N_G \approx 200$ 
(depending on the details of the identification procedure, 
Sect.~\ref{sez:identification}).
This is significantly larger than in most previous studies, 
where grouping procedures similar to ours were applied to
galaxy data of comparable quality to (but smaller extent than) PPS.
To avoid possible confusion, we note explicitly
that here we deal with {\emph loose} groups of galaxies.
Many studies concentrated on {\emph compact} groups,
a rather special case of the more general loose groups
considered here. 
The relation among compact and loose groups is discussed in 
Diaferio et al.(1994), Mamon (1996a), Governato et al. (1996).

Our group catalog is meant to be as homogeneous as possible
to those previously published and well--studied.
Unfortunately, group properties are very sensitive to the details
of the identification recipe (e.g., Pisani et al. 1992).
On the other hand, 
most of the actually available samples of loose groups 
were compiled following the same grouping criteria,
the adaptive Friends--Of--Friends (FOF) algorithms
introduced by Huchra \& Geller (1982, HG82 hereafter;
see also Nolthenius \& White 1987, NW87 hereafter).
Still, group properties are systematically influenced by
the user's choice of search parameters 
(HG82; 
NW87;
Moore et al. 1993;
Nolthenius et al. 1994, 1997; 
Frederic 1995a, 1995b; 
MFW93, NKP94, NKP97, F95a, F95b hereafter).
This must be taken into account by a careful, 
self--consistent match of FOF parameters among different catalogs.
Several authors (NW87; MFW93; NKP94; NKP97; F95a,~b)
used cosmological N--body simulations of dark matter models
to calibrate the ``optimal'' FOF algorithm.
In the first place, this aims at obtaining 
the highest possible completeness and reliability of FOF groups
(NW87; MFW93; F95a,~b).
It is worth to mention that this approach can then be reverted
in order to constrain the models. 
Once a grouping procedure is found successful, 
it is applied to real and simulated data, 
and outputs are self--consistently compared (NW87; NKP94; NKP97).

The earliest sample of FOF groups is the HG82 catalog ($N_G=92$)
based on the NB survey of nearby bright galaxies.
Geller \& Huchra (1983) and Nolthenius (1993; N93 hereafter)
compiled two larger catalogs ($N_G \sim 170$) 
from the CfA1 survey (Huchra et al. 1983).
These and similar CfA1 samples have been widely studied
(Mezzetti et al. 1985; Heisler et al. 1985;
Giuricin et al. 1986a,~b, 1988;
NW87;
Pisani et al. 1992;
MFW93;
N93;
NKP94, NKP97).
Groups were then selected from deeper galaxy surveys.
Maia et al. (1989; Maia \& da Costa 1990) 
identified $N_G=87$ groups in the SSRS1 survey
(da Costa et al. 1988),
while Ramella et al. (1989; RGH89 hereafter)
selected $N_G \sim 130$ systems from the CfA2 Slices 
(de Lapparent et al. 1986; Huchra et al. 1990, 1995).
Groups in the Las Campanas Redshift Survey (Shectman et al. 1996)
were considered by Tucker et al. (1993) and by Tucker (1994) in his Ph.D. Thesis. 
Their study is still underway (Tucker et al. 1997).

Very recently Ramella et al. (1997a; RPG97 herefter) 
published a larger group sample ($N_G=406$),
previously  announced by Pisani et al. (1994; PGHR94),
based on the whole CfA2 North survey.
RPG97 also announced the compilation of a further group catalog
whose details should be soon provided (Ramella et al. 1997b).
It should include the SSRS2 survey (da Costa et al. 1994),
which lies in the Southern Galactic Hemisphere as PPS
but in a completely independent area of the sky.
Loose groups in PPS ($N_G \sim 200$)
were systematically identified and analyzed in
Trasarti--Battistoni (1996; TB96 hereafter) in his Ph.D. Thesis, and
Trasarti--Battistoni et al. (1997; TBIB97 hereafter).
Two earlier studies are due to
Haynes \& Giovanelli (1988) and Wegner et al. (1993).
There, 
groups were selected from a much smaller subsample of PPS than we do here,
and they were mainly considereded as useful tracers of LSS.

Up to date, the largest group catalog ($N_G = 453$) is that of Garcia (1993).
On the other hand, 
the parent galaxy catalog EDB (Garcia et al. 1993)
is not a homogeneous redshift survey, but rather it is based on 
a compilation of galaxy data coming from very different sources,
though a great effort toward homogeneization 
of galaxy data (Paturel et al. 1989a,~b) was actually made.
The sample depth is  $B=14.0$, much shallower than for PPS, CfA, or SSRS.
Furthermore, 
groups are identified by means of FOF algorithms as well as other techniques.
This precludes any direct, straightforward comparison of Garcia's groups
with most available group samples and numerical simulations.
Similar criticisms apply to the group catalogs identified from the PGC sample
(Gourgoulhon et al. 1992; Fouqu\'e et al. 1992), and 
to a lesser extent to other group catalogs (Tully 1987; Giudice 1995, 1997).

The PPS sample is ideal for our purposes.
It is highly homogeneous,
apparent--magnitude--complete, and covers a wide solid angle. 
Furthermore,
it is based on the same parent angular catalog CGCG (Zwicky et al. 1961--68)
as the CfA survey,
but is deeper than CfA1 and SSRS1, wider than the CfA2 Slices,
contains more galaxies than each of such samples,
and it is directly comparable to the CfA2 North and SSRS2 surveys.
Groups are identified with the FOF recipe of HG82, 
but our search parameters match those adopted for the other group catalogs
compiled from galaxy samples of the same depth as ours.
Thus, 
our catalogs can be {\emph directly} compared--with/combined--to 
other
observational samples and/or numerical simulations of cosmological models
(in particular: RGH89, RPG97, F95a,~b).
In fact, this approach already allowed us (TBIB97) 
to compare group clustering in PPS with previous analyses 
of CfA1, SSRS1, and CfA2 Slices 
(Jing \& Zhang 1988; Maia et al. 1989; Ramella et al. 1990).
There, we show 
that many previously unexplained discrepancies among such analysis 
are essentially due to the different FOF parameters adopted 
by different authors.
This clearly indicates the need of a careful choice 
of search parameters prior to any comparison of different group samples.

The plan of the paper is the following. 
We describe the galaxy data in
Section \ref{sez:data},
and the group identification procedure in  
Section \ref{sez:identification}. 
The catalogs of groups and group properties are presented in
Section \ref{sez:results}.
Section \ref{sez:conclusion} is a summary.
Distances are measured in $\hMpc$, where the Hubble parameter is
$H_0 = 100 \ h \ \kms \ \Mpc^{-1}$,
and absolute magnitudes are computed assuming $h=1$.

\begin{figure*}
\begin{center} 
\epsfxsize=12cm
\begin{minipage}{\epsfxsize}\epsffile{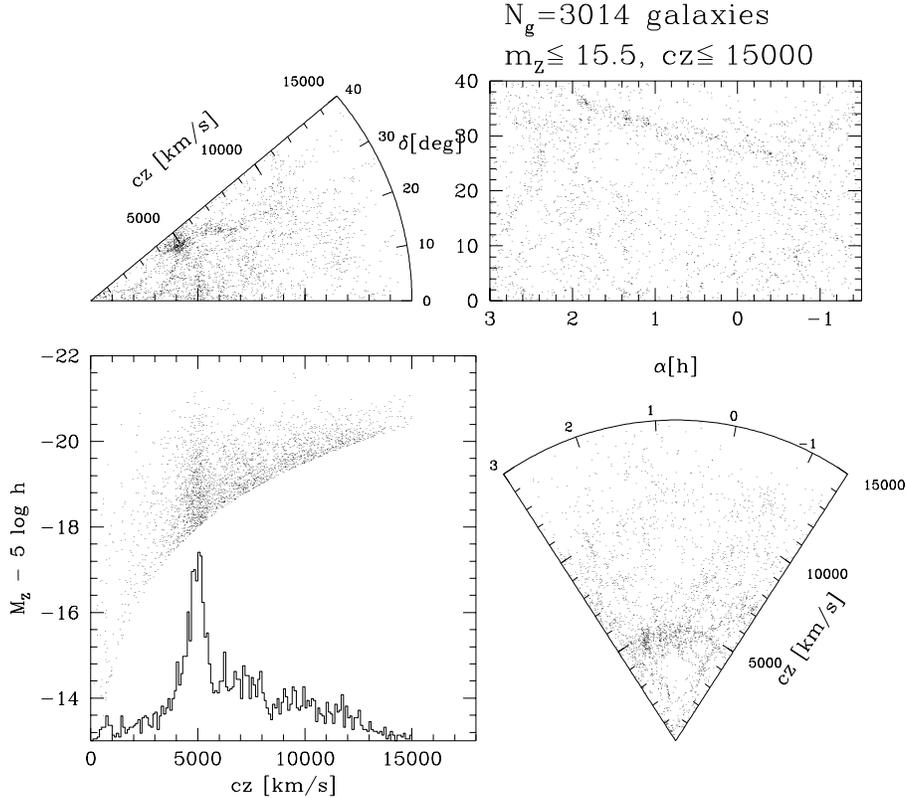}\end{minipage}
\end{center} 
\caption{\label{fig1gal}
Galaxies in the sample PPS2: 4+1 diagram, i.e. 4 sections of
$(\alpha, \delta, cz; M_Z)$ 
space + 1 redshift distribution histogram.
Each dot is a galaxy. The redshift scale is the same everywhere. 
Top left:  redshift space, $\delta$--$cz$ wedge diagram.
Top right: sky--view, $\alpha$--$\delta$ diagram.
Bottom right: redshift space, $\alpha$--$cz$ wedge diagram.
Bottom left: distance--luminosity diagram in the $cz$--$M_Z$ plane
(the lower envelope is the faint cut--off $\mlim=15.5$),
and distance (redshift) distribution histogram
(the normalization is arbitrary).}
\end{figure*}

\section{Galaxy Data}
\label{sez:data}

The PPS database was compiled by Giovanelli \& Haynes in the last decade 
(Haynes \& Giovanelli 1988; Giovanelli \& Haynes 1991; 
and references therein). 
It consists mainly of highly accurate 21--cm HI line redshifts, 
partly unpublished, obtained with the NAIC 305--m telescope in Arecibo and with
the NRAO 300--foot telescope formerly in Green Bank
(Giovanelli \& Haynes 1993, and earlier references therein).
The radio data are complemented with optical observations of early--type 
galaxies carried out at the 2.4--m telescope of the McGraw--Hill Observatory
(Wegner et al. 1993). 
The full redshift survey is magnitude--limited down to $m_Z \leq 15.7$,
and now it covers the whole region 
$-2^h.00 \le \alpha \le +4^h.00$ and $ 0^o    \le \delta \le 50^o   $.

From PPS, we extract a subsample named PPS2
(in analogy with CfA2),
complete and magnitude--limited to $m_Z \leq 15.5$.
We restrict PPS2 to the region 
$-1^h.50 \le \alpha \le +3^h.00$ and $ 0^o    \le \delta \le 40^o   $,
in order to exclude those parts near the northern edge of PPS
heavily affected by Milky Way obscuration.
Before the final selection,
we also correct Zwicky magnitudes (Zwicky et al. 1961--68)
for interstellar extinction as in Burstein \& Heiles (1978),
yielding $\delta m \leq 0.4$ over most of the selected area 
(see Fig.~2 in Giovanelli et al. 1986, or Fig.~1 in Park et al. 1994).
The solid angle is $\omega= 0.76 \ {\rm sr}$, 
and the degree of completeness is higher than 95\% to this magnitude limit 
(see Fig.~1 in Iovino et al. 1993, 
who used a similar but earlier version of PPS).

Regarding redshifts, we consider three different schemes:
(a) a correction of observed heliocentric radial velocities 
as in Yahil et al. (1977),
i.e. 
taking into account the motion of the Sun 
relative to the rest frame of the Local Group Centroid,
$v_{\odot LGC}=308 \ \kms$ towards $l=105^o$, $b=-7^o$;
(b) a correction of $cz$ for the motion of the Sun relative to
the rest frame of the Microwave Background Radiation,
$v_{\odot MBR}=270 \ \kms$ towards $l=265^o$, $b=+48^o$ (e.g. Peebles 1993);
(c) no correction at all.
The total number of galaxies 
is very little affected by purely radial corrections to $cz$.
In complete apparent--magnitude--limited samples, 
galaxies ``flow'' along the lines of constant $m$
in the $cz$--$M$ plane, but only very few ``move across'' the redshift border.
(This effect would not be negligible in complete volume--limited samples, 
i.e all galaxies within a certain range of $cz$ and absolute magnitude $M$,
where corrections to $cz$ cause ``flows'' in the $cz$--$M$ plane that
cross both the redshift and the luminosity edges of the selected sample.)
We expect group properties to be very weakly affected too, 
as all member galaxies within a given group receive similar corrections
(they are ``moved all together'' in redshift space, 
and group centroids ``follow'' them).

Figure~\ref{fig1gal} shows PPS2 in redshift+luminosity space.
This ``4+1--diagram'' (i.e., 3 maps in redshift space and 
1 luminosity--distance diagram +1 redshift distribution histogram),
eases comparison of group catalogs among them and 
with their parent galaxy sample.
The final sample PPS2 contains $N_g=3014$ galaxies, 
with extinction--corrected magnitude $m_Z \leq 15.5$ and
MBR--corrected redshift  $cz  \le 27000 \ \kms$ 
(in practice, almost all galaxies are contained within $cz \sim 15000 \ \kms$).

For the sake of comparison, the samples 
CfA1 (North + South) and the (first two, northern) CfA2 Slices 
are characterized by $\omega=1.83+0.83$ and $0.42 \ {\rm sr}$, 
$m_Z \leq 14.5$ and $15.5$, $N_g=1845+556$ and $1766$ respectively.
The SSRS1 survey is apparent--diameter--limited, 
with $N_g=2028$, $\omega=1.75 \ {\rm sr}$, and
$m_Z \simlt 14.8$, slightly deeper than CfA1.
The SSRS2 and CfA2 North surveys have
$N_g \sim 3600$ and $6000$, $\omega=1.13$ and $1.2 \ {\rm sr}$, 
respectively, and $m_Z \leq 15.5$.
We note that magnitudes in the CfA samples are not corrected 
for galactic extinctions, and redshifts are usually corrected 
for the solar motion with respect to the Local Group
(sometimes also for infall on the Virgo cluster).

\section{Group identification procedure}
\label{sez:identification}

In early catalogs, 
galaxy groups were identified ``by eye'' (e.g., de Vacouleurs 1975)
and/or in projection (e.g., Turner \& Gott 1976) from angular galaxy catalogs.
Nowadays, groups are better identified by means of
objective grouping procedures applied to galaxy redshift surveys.
Several such prescriptions have been suggested in the literature
(Turner \& Gott 1976; Materne 1978, 1979; Paturel 1979; 
Tully 1980, 1987; HG82; NW87; N93; Pisani 1993, 1996).

\subsection{Friends--Of--Friends algorithm}
\label{sez:FOF}

We adopt the adaptive FOF algorithms introduced by HG82 for several reasons.
First, most loose group catalogs extracted from galaxy redshift surveys
(HG82; Geller \& Huchra 1983; RGH89; Maia et al. 1989; N93; Garcia 1993; PGHR94; RPG97)
are based on this technique, and we want to compare them with ours.
Second, FOF algorithms are relatively faster and easier to implement than 
other objective grouping algorithms
(Turner \& Gott 1976; Materne 1978, 1979; Paturel 1979; 
Tully 1980, 1987; Pisani 1993, 1996),
lead to a unique output catalog for given input parameters,
and do not rely on any a priori
assumption regard to the geometrical shape of galaxy groups.
Third, FOF algorithms are well--studied tools, as 
they have been repeatedly applied to numerical simulations of galaxy surveys.
These have been either
(i) cosmological N--body simulations 
(NW87; MFW93; NKP94, NKP97; F95a,~b),
where all six space and velocity coordinates are known in advance
and allow for a self--consistent matching of ``real'' and FOF groups, or
(ii) geometrical Monte--Carlo simulations (RGH89; RPG97)
which accurately mimick the main LSS features of a given data set
and allow to assess the impact of LSS on the group properties.
All such simulations provided evidence that FOF--identified objects
indeed mostly correspond to physically real galaxy groups, 
though poor groups with only $N_{mem}=3$ or $4$ members may be
substantially contaminated (RGH89; F95a; RPG97).
No similar extensive countercheck on the other group--finding techniques
has been reported yet.
Fourth, further and more direct evidence of the ``reality'' of FOF groups
was recently provided by direct observation of the 
neighbourhood of FOF groups by Ramella et al. (1995a--b, 1996), who also showed how
the physical properties of FOF--identified loose groups are a reliable estimate
of those of the ``real'' underlying galaxy groups.
To be fair, it should be noted that the HG82 recipe tends to include 
a high fraction of spurious members and/or groups (NW87; F95a,~b).
However, the compilation of a catalog requires, 
in the first place, a high degree of completeness;
suspicious objects can still be discarded later on.

The operational definition of a group is 
a number density enhancement in (redshift) space 
(HG82; NW87 have a slightly different point of view). 
The algorithm can be thought of as a percolation technique, 
but truncated to a specified value $R_0$ of the connecting link $R_L$:
galaxies closer than $R_0$ are ``friends'' of each other, 
friendship is transitive, 
and an isolated set of (at least 3) friends is what we call a galaxy group.
In the ideal case of a luminosity--complete and volume--limited sample with
$N$ galaxies,  volume $V$, average density $\bar n = N/V$, 
with no redshift distortions,
groups would then be selected above a fixed number density threshold 
$\delta n/n$, given by:
\begin{equation}
\label{eq:R0dn/n}
1 + {\delta n \over n}  = 
{3\over 4 \pi R_0^3 \bar n} =
{3 / \left ( 4 \pi R_0^3\right ) \over 
\int_{-\infty}^{\Mlim} \phi (M) dM } \ .
\end{equation}
where $\phi(M)$ is the galaxy LF of the sample 
[Sect.~\ref{sez:LFSF}, Eq.~(\ref{eq:LF})] and all galaxies are 
brighter than the absolute magnitude completeness limit $\Mlim$.

\subsection{Radial scaling of the links}
\label{sez:scaling}

In practice, loose groups have usually been identified from 
apparent--magnitude--limited redshift surveys.
This brings in two main complications:
(i) strong radial {\emph redshift distortions} due to peculiar motions,
mainly induced by small--scale galaxy dynamics within the groups themselves, 
and
(ii) distance {\emph selection effects} 
due to the difficulty of observing fainter galaxies at larger distances.
Neglecting the effect of strong spatial inhomogeneities (LSS),
the expected number density of galaxies at a distance $r$ from us, 
brighter than $\mlim$ in apparent magnitude and brighter than 
$\Mlim(r)= \mlim -2.5\log_{10}(hr/\Mpc) -25$ in absolute magnitude,
is given by
\begin{equation}
\label{eq:nrphi}
\bar n(r;\mlim) = \int_{-\infty}^{\Mlim (r)} \phi (M) dM \ 
\end{equation}
which increases with $\mlim$ and decreases with $r$.

To overcome effect (ii),
the strategy is to ``compensate'' the decrease in $\bar n(r; \mlim)$
by allowing the links to be ``more generous'' at larger $r$ or fainter $\mlim$.
To deal with effect (i),
the spatial link $R_L$ is replaced by
a transverse ``sky--link'' $D_L$ and a radial ``redshift--link'' $V_L$.
Different authors proposed qualitatively different solutions
to implement such a strategy
(HG82; NW87; see also Gourgoulhon et al. 1992, N93, Garcia 1993;
we refer to the original papers for details), 
involving also dynamical considerations about $V_L$ because of effect (i).
Advantages and shortcomings of one recipe over another are discussed in 
NW87, Garcia (1993) and F95a,~b.

We adopt the scaling recipe of HG82, though with stricter normalizations.
Both $D_L$ and $V_L$ are normalized by $D_0 \equiv D_L(cz_0)$ and $V_0 \equiv V_L(cz_0)$
at some fiducial redshift $cz_0=H_0 r_0$.
They are then scaled with $cz$ as $\left[ \bar n(cz_{ij}/H_0;\mlim) \right]^{-1/3}$,
where $cz_{ij}=(cz_i+cz_j)/2$ is the median redshift of the $ij$-th pair of galaxies.
Galaxies are linked if their transverse and radial separations 
$r_{ij}^{\perp}$ and $r_{ij}^{\parallel}$ satisfy
$r_{ij}^{\perp}     \leq D_L(cz_{ij})$ and 
$r_{ij}^{\parallel} \leq V_L(cz_{ij})/H_0$, respectively.
The number density within groups at distance $r$ is
\begin{equation}
\label{eq:ngrp}
n_{\rm grp} (r;\mlim) \geq {3\over 4 \pi D_L^3 (r)} = 
\left (1 + {\delta n \over n} \right ) 
\bar n(r;\mlim) \ ,
\end{equation}
which with the HG82 scaling $D_L(r) \propto \bar n(r)^{-1/3}$ yields
\begin{equation}
\label{eq:D0dn/n}
1 + {\delta n \over n}  = 
{3\over 4 \pi D_0^3 \bar n(r_0)} =
{3 / \left ( 4 \pi D_0^3\right ) \over 
\int_{-\infty}^{\Mlim (r_0)} \phi (M) dM } \ .
\end{equation}
We emphasize that the relation among $\delta n/n$ and $D_0$ 
depends on the galaxy LF.
Also, the spherical symmetry of the ``volume of friendship''
implicitly assumed in the idealized Eq.~(\ref{eq:R0dn/n}) 
is broken by the actual, 
anisotropic definition of friendship through the two links $D_0$ and $V_0$.

\subsection{Normalization of the links}
\label{sez:normalization}

As previously pointed out, group properties sensitively depend on the chosen algorithm.
On the other hand, we want to define group samples which may be 
directly compared--to/combined--with those previously published.
We discuss here how to face this problem.
We emphasize that here we concentrate on the question of how to ``match''
FOF algorithms for different data sets.
The complementary problem of how to calibrate the ``optimal'' (if any) FOF algorithm 
for a given data set has been extensively discussed by other studies 
(HG82; NW87; RGH89; MFW93; F95a, b; NKP94, NKP97), and it is out of our scope.

According to HG82, groups are selected above a density contrast
$\delta n/n$ {\emph given a priori}. 
The physical justification is the hypothesis--requirement that 
galaxy groups correspond to dynamically bound matter overdensities,
whose dynamical state is dictated by their density contrast $\delta \rho/\rho$,
in turn related to the number density contrast $\delta n/n$,
maybe through some mechanism of biased galaxy formation  (e.g., Kaiser 1984).
Unfortunately, one does not know exactly the value
of the normalization of $\phi(M)$ in Eq.~(\ref{eq:R0dn/n}).
Due to the presence of LSS, this may vary by a factor of 2 from survey to survey,
and fluctuates strongly even within the same galaxy survey
(eg., de Lapparent et al. 1988, 1989; see also Sect.~\ref{sez:LFSF}).
Fluctuations and uncertainties about the shape of the LF
also have a (slight) effect on the $D_0$--$\delta n/n$ relation~(\ref{eq:D0dn/n})
(TB96; Sect.~\ref{sez:variability}).
Moreover, the relation among 
the (physically well--motivated) {\emph mass} density--contrast $\delta \rho/\rho$ 
and the observatio {\emph galaxy number} density--contrast $\delta n/n$,
is still very uncertain (e.g., Bower et al.~1993).
Actually, the very existence of a universal value of $\bar n$, valid at any 
sufficiently large spatial scale, has been repeatedly questioned 
(e.g., Coleman \& Pietronero 1992; Baryshev et al. 1994)
and the debate on this point is still alive
(Davis 1996; Pietronero et al. 1997).

What normalizations should we adopt?
For the sake of clarity, let us first consider only
data samples of the {\emph same depth}, but different galaxy LF.
(We will discuss the effect of different depth,
and of further sample--to--sample variations 
due to local LSS features, later on.)
The RGH89 group catalog in the CfA2 Slices is normalized by
$V_0 = 350 \ \kms$ and $D_0 = 0.270 \ \hMpc$ at $cz_0=1000 \ \kms$.
With the galaxy LF adopted by RGH89, $D_0$ translates into $\delta n/n =80$.
This very value may be regarded as an update of $\delta n/n =20$
used by HG82 and based on theoretical considerations, 
enhanced by a factor 4 to account for the locally high density of the 
the Great Wall region within CfA2 Slice where most of the groups reside 
(Ramella et al. 1992; Ramella, private communication).
Subsequent observational (Ramella et al. 1995a-b, 1996) and theoretical (F95a,~b) studies 
confirmed such values of $V_0$ and $D_0$
as ``optimal'' for the compilation of a group catalog,
though different normalizations may be more appropriate in different contexts 
(NW87; MFW93; N93; F95a,~b; NKP94, NKP97).
We also note that the specific conclusions of most such works should be taken with caution, 
as (i) they do not always use the precise HG82 scaling for $D_L$ and $V_L$,
and (ii) their observational sample is often the CfA1 survey, 
brighter and shallower than CfA2 and PPS2. (For a discussion of the effect of depth, see below.)
Later, PGHR94 and RPG97 used a slightly different $D_0=0.231 \ \hMpc$, 
due to the higher density of CfA2 North and to their requirement $\delta n/n=80$.
The galaxy LF of PPS2 is still different, 
yielding further combinations of $D_0$ and $\delta n/n$.
With our choice of $\phi(M)$ (Sect.~\ref{sez:LFSF}), 
$D_0=0.231$ and $0.270 \ \hMpc$ yield $\delta n/n = 173$ and $108$, respectively,
instead of the desired $\delta n/n=80$.
The latter is only recovered if we adopt $D_0=0.300 \ \hMpc$,
substantially larger than $D_0$ used by RPG97.
This is consistent with the different normalizations
suitable for PPS2, CfA2 Slice, and CfA2 North
($\phi_*=0.02 \pm 0.1 \ h^3 \ \Mpc^{-3}$,
$\phi_*=0.025 \ h^3 \ \Mpc^{-3}$, and
$\phi_*=0.05 \pm 0.2 \ h^3 \ \Mpc^{-3}$, respectively)
and well within the LF normalization uncertainty.

We are then led to the following question.
In order to build group catalogs physically as similar as possible to each other,
should they be compiled using (the same link $V_0$ and)
the {\emph same link} $D_0$, or 
the {\emph same density threshold} $\delta n/n$?
We emphasize that the parameter $\delta n/n$ defined in Eq.~(\ref{eq:D0dn/n}) 
is customarily used only in order to {\emph label} a given FOF catalog.
In fact, no such assumption is needed to {\emph identify} the groups,
and one could as well use as label the $D_0$ parameter 
effectively used by the FOF algorithm itself.
Moreover, the theoretically motivated mass overdensity $\delta \rho/\rho$
is ideally referred to the mean mass density $\bar \rho_0$ of the whole Universe,
and not $\bar \rho_S$ averaged only over the considered sample. 
However, no matter whether the universal value of $\bar\rho_0$ 
coincides with $\bar \rho_S$ and whether it is known or not, 
for a given $\bar \rho_0$ identical $D_0$'s will correspond to identical $\delta \rho /\rho$.
So, though strong theoretical motivations suggest $\delta \rho / \rho$ to be the relevant
physical quantity, and suggest $\delta n/n$ as its observational counterpart,
on practical grounds it is more justified to relate the link normalization directly to $D_0$.
As a matter of fact, 
the issue of $D_0$ vs $\delta n/n$ was considered also by Maia et al. (1989).
They compared FOF groups in SSRS1 and in CfA1, identified either with the same $D_0$
or with the same $\delta n/n$, and found that the median physical properties
of the groups were more similar in the former case than in the latter.
However, it is not clear how to interpret such result, since
the two galaxy samples SSRS1 and CfA1 differ in depth and selection criteria.

We can now discuss the case of data samples of {\emph different depth},
for which things are still slightly more complicated.
For simplicity, we assume them to have the same parent LF.
We adopt the Schechter (1976) functional form
$\varphi(L/L_*)d(L/L_*)=\phi_* (L/L_*)^{-\alpha} \exp(-L/L_*)d(L/L_*)$.
Then from Eq.~(\ref{eq:D0dn/n}) above, constant $\delta n / n$ requires
\begin{equation}
\label{eq:D0gam}
D_0 \propto
\left [\int_{(r/r_*)^2}^\infty \varphi ({L\over L_*}) d ({L\over L_*}) \right
]^{-1/3} \!\!\!\!\!\!\!\!
\propto \left (\Gamma \left [1\!\!+\!\!\alpha, ({r\over r_*})^2\right ] \right )^{-1/3}
\end{equation}
where $r_*$ is the maximum distance where 
a galaxy of luminosity $L_*$ (absolute magnitude $M_*$) is still observable. 
Adopting $r_0 = 10 \ \hMpc$ ($cz_0 = 1000 \ \kms$), 
$M_* = -19.3 + 5 \log h$, and $\alpha = -1.15$,
going from a catalog limited at $\mlim=14.5$ to one limited at $\mlim=15.5$, 
$r_*$ goes from 58 to $91 \ \hMpc$, the minimum $L/L_*$ from $(r_0/r_*)^2=0.030$ to 0.012, 
so to keep $\delta n/n$ constant at the normalization location $r_0$ 
the link normalizations should be related by 
\begin{equation}
\label{eq:D0D0}
{D_0 (\mlim=14.5) \over D_0 (\mlim=15.5)}=
\left [
{\Gamma (-0.15, 0.030) \over \Gamma (-0.15, 0.012) }
\right ]^{-1/3}
= 1.12 \ .
\end{equation}
In other words, at any given distance $r$ the deeper sample 
contains more galaxies and has a higher average density $\bar n(r;\mlim)$, 
so fixing the same $\delta n/n$ at a given $r_0$ yields 
a smaller $D_0$ for fainter $\mlim$, albeit only by $12 \%$.
One way out of this technical difficulty (TB96)
is to choose a normalization location $\tilde r_0(\mlim)$ 
variable from sample to sample, and equal to 
(i) a constant fraction of the characteristic sample depth $r_*$, or
(ii) zero distance. The problem with  
(i) is to introduce a dependence on $L_*$ (which also varies from sample to sample),
while (ii) is technically delicate as Eq.~(\ref{eq:nrphi}) with $\phi$ given 
by Eq.~(\ref{eq:LF}) formally diverges for $r \rightarrow 0$, 
but suitable limits can still be defined (TB96).
In practice, given the depth of the galaxy samples from which galaxy groups can be 
meaningfully identified, $cz_0 = 1000 \ \kms$ is already close enough to zero that 
simply keeping the same normalization location $r_0$ for samples of different $\mlim$
introduces only a small inconsistency in the values $D_0$ and $V_0$ 
[Eq.~(\ref{eq:D0D0})], largely overwhelmed by the other sources of uncertainty.

All these complications, due to the nature of the FOF grouping algorithm,
would be avoided if group properties turned out
to change little for reasonable variations of FOF parameters. 
Actually, RPG97 report that group properties are non--significantly sensitive 
to the choice of linking parameters over a wide range of $\delta n/n$.
On the other hand, the symmetry and the ``orthogonality'' of $D_L$ and $V_L$ 
in the HG82 algorithm suggests
$D_0$ to be directly related to the group size ($R_{\rm h}$ or $R_p$), and 
$V_0$ to be directly related to the group velocity dispersion $\sigma_{v}$,
possibly with some residual ``non--orthogonal'' dependence.
Since $(1+\delta n/n) \propto D_0^{-3}$, 
any dependence of a given group property $X$ on $D_0$ or $\delta n/n$
is equally recovered if one plots $X$ against $\log D_0$ or $\log(1+\delta n/n)$.
In summary, even a substantial dependence of group properties on the links $V_0$ and $D_0$, 
though not removed, might be missed or hidden 
by focusing attention only on the customary $\delta n/n$ parametrization.

\begin{figure*}
\begin{center} 
\epsfxsize=12cm
\begin{minipage}{\epsfxsize}\epsffile[18 420 580 700]{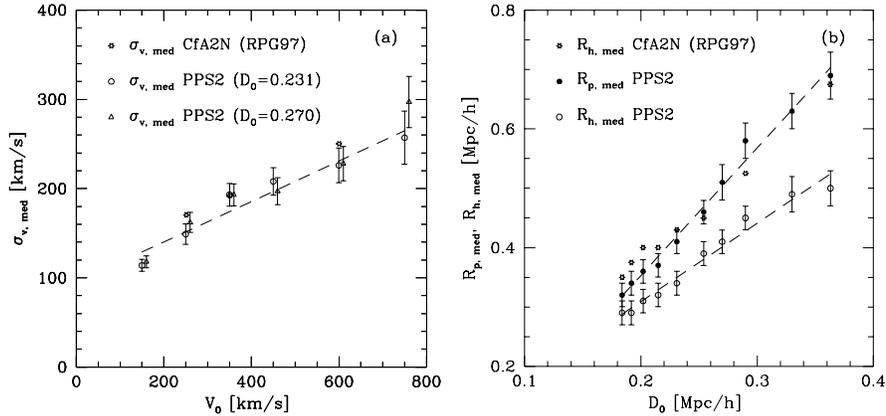}\end{minipage} 
\end{center} 
\caption{\label{fig2medians}
Dependence of group properties (medians) from the FOF normalizations.
(a) the line--of--sight velocity dispersion $\sigma_v$ vs the redshift link $V_0$ ($D_0=0.231 \ \hMpc$);
(b) the harmonic mean radius $R_{\rm h}$, and 
the pairwise mean separation $R_{\rm p}$ vs the spatial link $D_0$ ($V_0=350 \ \kms$).
Symbols are explained in the Figure.
The straight lines are linearly fitted to the data of the PPS2 groups only.
The bars are $\pm 1$ standard deviation divided by $\sqrt{N_G}$.
Such bars have not been used in the linear fit to the medians.
They are only shown in order to give an idea of the typical dispersions
of group properties around their central values.
We also show data for the CfA2N groups of Ramella et al. (1997; their Fig.~1),
and for PPS2 groups selected with $D_0=0.270 \ \hMpc$ and variable $V_0$.}
\end{figure*}

In order to test such effects on group properties,
we built several arrays of group catalogs with different normalizations,
and the same LF adopted for PPS2.
The first array has given $V_0=350 \ \kms$ and variable 
$D_0=$
$0.184$, $0.192$, $0.202$, $0.215$, $0.231$, $0.254$, 
$0.270$, $0.290$, $0.300$, $0.330$, $0.363 \ \hMpc$, 
including the $D_0$'s used by RPG97 (Ramella, private communication) and yielding
$\delta n/n=$
$342$, $301$, $259$, $214$, $173$, $129$, 
$108$, $87$, $79$, $59$, $45$, respectively, with our LF.
The second array has given $D_0=0.231 \ \hMpc$ and variable 
$V_0= 150$, $250$, $350$, $450$, $600$, $750 \ \kms$.
The third array is like the second one, but for $D_0=0.270 \ \hMpc$.
In Figure~\ref{fig2medians}a, we plot 
the line--of--sight velocity dispersion $\sigma_{v}$ (medians) against the redshift link $V_0$.
Analogously, 
in Figure~\ref{fig2medians}b, we plot
the mean harmonic radius $R_{\rm h}$ and mean pairwise member separation $R_p$ (medians)
against the spatial link $D_0$.
In all cases, we find strong, approximately linear correlations:
$\sigma_{\rm v, med} \simeq 0.23 V_0 + 95 \ \kms$,
$R_{\rm h, med} \simeq 1.32 D_0 +0.04 \ \hMpc$,
$R_{\rm p, med} \simeq 2.15 D_0 -0.08 \ \hMpc$.
Similar results also hold for groups 
in CfA2 North (Fig.~\ref{fig2medians}, RPG97's data), in CfA1 (NKP94; NKP97), 
and in SSRS1 (Maia et al. 1989, their Table 5).
Although RPG97 produce a figure with a decreasing trend of $R_{\rm h}$
versus $delta n /over n$, they do not mention that this trend
simply arises from the proportionality between median $R_{\rm h}$ and $D_0$
(which would have been obvious had they made a log--log plot).

For all the above mentioned reasons, several of which are mainly practical, 
we prefer to parametrize a given FOF algorithm by its $D_0$ and $V_0$. 
In particular, we specify the threshold spatial separation used by the FOF algorithm 
{\emph directly} in terms of $D_0$, instead of through the density contrast (HG82)
or as a fraction of the mean inter galaxy separation (NW87).
We then compute $\delta n/n$ {\emph a posteriori} through Eq.~(\ref{eq:D0dn/n}),
and interpret its value mainly as an approximate measure,
rather than a precise parametrization, of density contrast.

\begin{table*}
\begin{center}
\caption{\label{tab_catgrp} The catalog loose groups in PPS2: two example lines}
\begin{tabular}{r r   r r   r r r r r r r r}
\hline 
    &   & &     & & & & & & & & \\
$i_G$  &$N_{mem}$ &$\alpha_{1950}$  &$\delta_{1950}$  
&$cz$  &$\sigma_v$ &$R_h$ &$R_p$ &$\log(L_G)$ &$\log(\Mvir)$ &$\log({\Mvir \over L_G})$ 
&$t_{cr}$ \\
    &  &hh mm.f &dd mm.f &$\kms$	&$\kms$	  &$\hMpc$ &$\hMpc$ & & & &$H_0^{-1}$	\\ 
    &   & &     & & & & & & & & \\
\hline
    &  &        &        &     &   &    &    &      &      &     &      \\
  1 &3 &-1 24.6 &33 34.5 &756  &40 &.24 &.24 &10.00 &11.74 &1.74 &.670  \\
188 &3 &2  55.8 &3  21.2 &2925 &19 &.24 &.22 &9.97  &11.08 &1.11 &1.324 \\
    &  &        &        &     &   &    &    &      &      &     &      \\
\hline
\end{tabular}
\end{center}
\end{table*}

\subsection{Luminosity Function}
\label{sez:LFSF}

The galaxy LF is parametrized with the Schechter (1976) form:
\begin{equation}
\label{eq:LF}
\phi(M)= const \cdot 
\left[10^{0.4(M_* - M)}\right]^{1+\alpha}
\exp\left[-10^{0.4(M_* - M)}\right]
\end{equation}
where $const = \phi_* \cdot 0.4 \cdot \ln 10$.
We assume distance proportional to $cz$,
extinction--corrected $m_Z$,
and no K--corrections (very small for this low redshift sample),

Recently, Marzke et al. (1994; MHG94 hereafter) estimated the LF of the 
whole CfA2 Survey, and several subsamples therein.
For their CfA2 South subsample, very similar to PPS2, they get: 
$\alpha =  -0.9 \pm 0.2$, 
$M_*   = -18.9 \pm 0.1$, 
$\phi_*=  0.02 \pm 0.01 \ h^3 \ \Mpc^{-3}$.
Note that they explicitly corrected for (a form of) Malmquist bias,
the Eddington bias:
Zwicky magnitudes have an uncertainty $\sigma_m \simeq 0.3$--$0.4$ mag, 
which causes a ``random diffusion'' of the more numerous fainter galaxies 
towards brighter magnitudes, and in turn modifies the overall shape of the LF.
This effect induces
an artificially bright $M_*$, and a correspondingly too negative $\alpha$.

We use an estimate of $\phi(M)$ based directly on our data (TB96).
First, though PPS2 and CfA2 South are similar samples, 
they are still slightly different.
Their galaxy LF's might be different, and this might effect
group identification through the $\phi(M)$ ingredient of the FOF algorithm.
Second, group identification requires a Malmquist--uncorrected $\phi(M)$. 
Malmquist corrections to $\phi(M)$ are global,
and they do not apply to each single galaxy.
In fact,
only the study of intrinsic physical properties of the galaxies themselves 
would require such (unknown) corrections 
(and others, e.g. internal extinction).
Our sample PPS2 is spatially inhomogeneous (Fig.~\ref{fig1gal}).
This requires a density--inhomogeneity--independent technique
(e.g., Efstathiou et al. 1988; de Lapparent et al. 1989; 
see also the review of Binggeli et al. 1988).

Using the inhomogeneity--independent STY method (Sandage, Tamman, \& Yahil 1979)
and Zwicky magnitudes (corrected for Milky--Way extinction),
TB96 found $\alpha = -1.15 \pm 0.15$ and $M_* = -19.3 \pm 0.1$.
By construction, the STY technique does not provide an estimate of the
density normalization $\phi_*$.
Simple tests with the non--parametric inhomogeneity--independent C--method
(Lynden--Bell 1971; Choloniewsky 1986,1987)
or, on the other hand, a countercheck with
the more na\"{\i}ve inhomogeneity dependent $1/V_{\rm MAX}$ technique 
(e.g., Felten 1977),
all seem to suggest $\phi_* \sim 0.01 \ h^3 \ \Mpc^{-3}$ for PPS2 (TB96).
However, matching the absolute magnitude counts $dN/dM$ or the radial counts $dN/dr$
is better accomplished using $\phi_* \sim 0.02$ (TB96).
Both values are consistent with the typical uncertainty $\Delta \phi_*/\phi_* \approx 0.5$,
and with the CfA2 South value $\phi_* =0.02 \pm 0.01 \ h^3 \ \Mpc^{-3}$ given by MHG94.
Since their analysis is superior TB96's on this point,
here we also adopt $\phi_* =0.02 \pm 0.01 \ h^3 \ \Mpc^{-3}$.
(Note that this only matters when we insist on translating
a given $D_0$ in the grouping algorithm into
an approximate density threshold $\delta n/n$.)

Regarding the other two Schechter parameters, 
estimating errorbars as in Marshall (1985) the same tests also suggested an uncertainty
$\Delta \alpha \approx \pm 0.15$ and $\Delta M_* \approx \pm 0.1$ (TB96).
This is close to the typical observational uncertainties 
(and/or the scatter among different techniques and/or data sets)
$\Delta M_*         \sim  0.1$-$0.2$ and
$\Delta \alpha      \sim  0.1$-$0.2$
reported by MHG94, and the similar earlier studies quoted above.
The results of TB96 (not Malmquist--corrected) and of MHG94
(Malmquist--corrected
by $\delta \alpha \simeq 0.1$-$0.2$, $\delta M_* \simeq 0.3$-$0.4$, both positive)
are in good agreement, once we take into account 
the different details between the two analyses.
In fact, by using inhomogeneity--independent techniques, 
several authors estimated uncorrected Schechter parameters and
their corresponding additive corrections $\delta M_*$ and $\delta \alpha$.
In the first two Northern CfA2 Slices, de Lapparent et al. (1989) found
$\delta M_* \simeq 0.3$ and $\delta \alpha \simeq 0.1$.
For the whole CfA1, similar values were found 
by Efstathiou et al. 
(1988;  $\delta M_*\simeq 0.39$ and $\delta \alpha \simeq 0.18$), and
by N93 ($\delta M_*\simeq 0.45$ and $\delta \alpha \simeq 0.24$).
Our LF and those of de Lapparent et al. (1988, 1989) for the CfA2 Slice(s)
($\alpha=-1.2$-$1.1 \pm 0.1$, $M_*=-19.15$-$19.2 \pm 0.1$, 
not Malmquist--corrected),
used in previous group catalogs, are also rather similar.
Let us outline that, concerning group properties, 
such residual small differences 
have however a rather small effect (Sect.~\ref{sez:variability}; TB96).

\begin{figure*}
\begin{center} 
\epsfxsize=12cm
\begin{minipage}{\epsfxsize}\epsffile{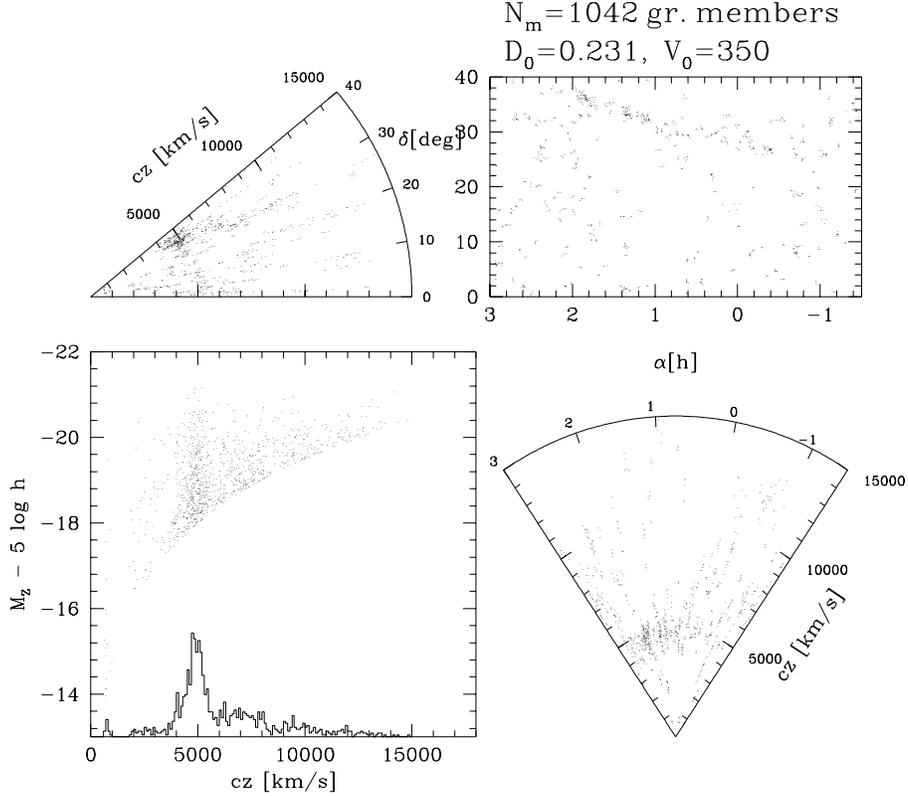}\end{minipage} 
\end{center} 
\caption{\label{fig3mem}
Members of loose groups in PPS2: 4+1 diagram.
Each dot is a group member galaxy. Everything else as in Fig.~1.}
\end{figure*}

\begin{figure*}
\begin{center} 
\epsfxsize=12cm
\begin{minipage}{\epsfxsize}\epsffile{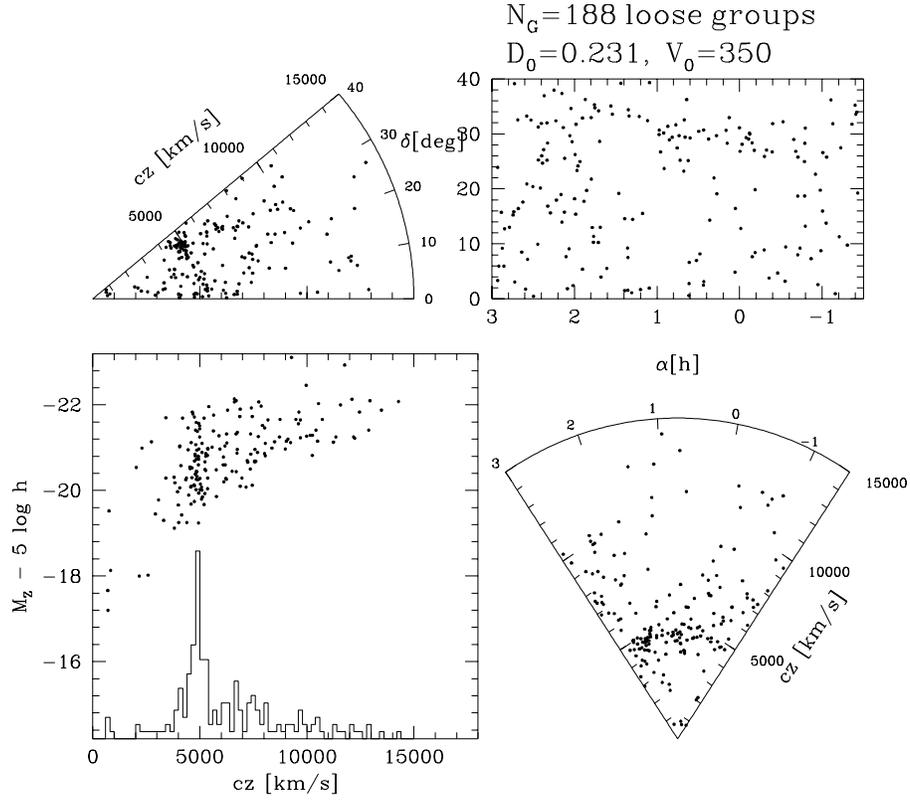}\end{minipage} 
\end{center} 
\caption{\label{fig4grp} Loose Groups in PPS2: 4+1 diagram.
Each dot is a group
(number--weighted centroid, $N_{mem}\geq 3$ member galaxies).
The group magnitude is computed by
adding up the luminosity of all observed group members,
and the lower envelope corresponds to 3 member galaxies 
of magnitude $\mlim=15.5$. Everything else as in Fig.~1.
}
\end{figure*}

\begin{figure*}
\begin{center} 
\epsfxsize=12cm
\begin{minipage}{\epsfxsize}\epsffile[40 410 570 700]{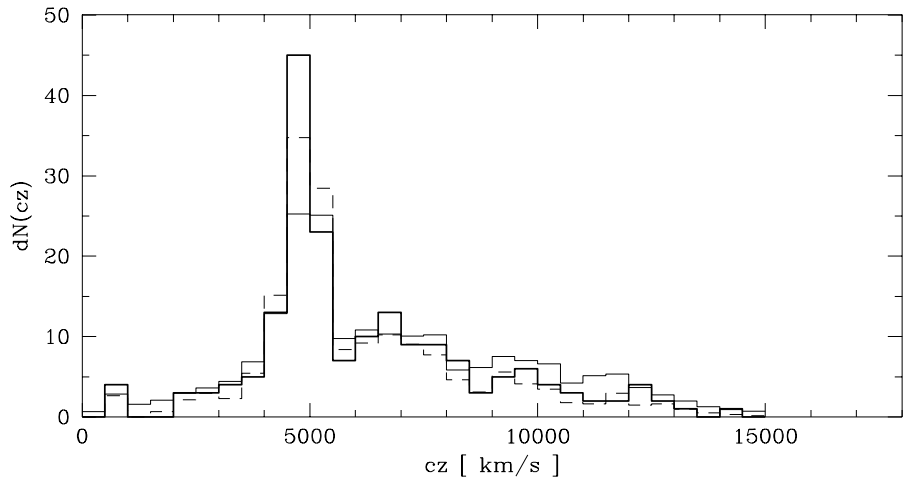}\end{minipage} 
\end{center} 
\caption{\label{fig5histo}
Radial distribution of groups and galaxies in PPS2.
Redshift distribution for 
groups (thick histogram), 
group members (thin dashed histogram), and 
all  galaxies (thin dotted histogram).
For comparison, all curves are normalized to the same total area
}
\end{figure*}

\begin{figure*}
\begin{center} 
\epsfxsize=12cm
\begin{minipage}{\epsfxsize}\epsffile{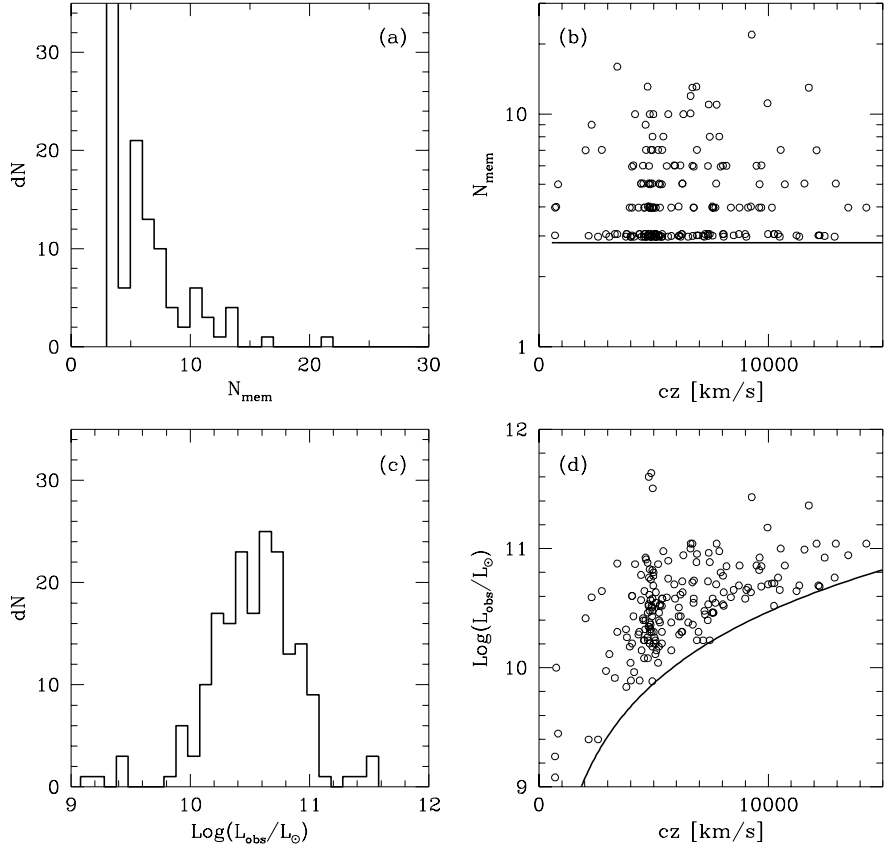}\end{minipage} 
\end{center} 
\caption{\label{fig6abcd}
Internal properties $X$ of loose groups in PPS2.
Left panels: distribution histogram $dN(X)$.
Right panels: variation with distance.
Here we have: 
(a,b) $X=N_{mem}$;
(c,d) $X=\log_{10}(L_G/L_\odot)$.
The curves are not fitted to the data.
Here, the horizontal solid line corresponds to $N_{mem}=3$
and the smooth curve is the magnitude limit for groups,
corresponding to $3 L_{\rm lim}(cz;\mlim=15.5)$ for galaxies.
We assume $h=1.0$ and $M_\odot=+5.48$ (blue).
}
\end{figure*}

\begin{figure*}
\begin{center} 
\epsfxsize=12cm
\begin{minipage}{\epsfxsize}\epsffile{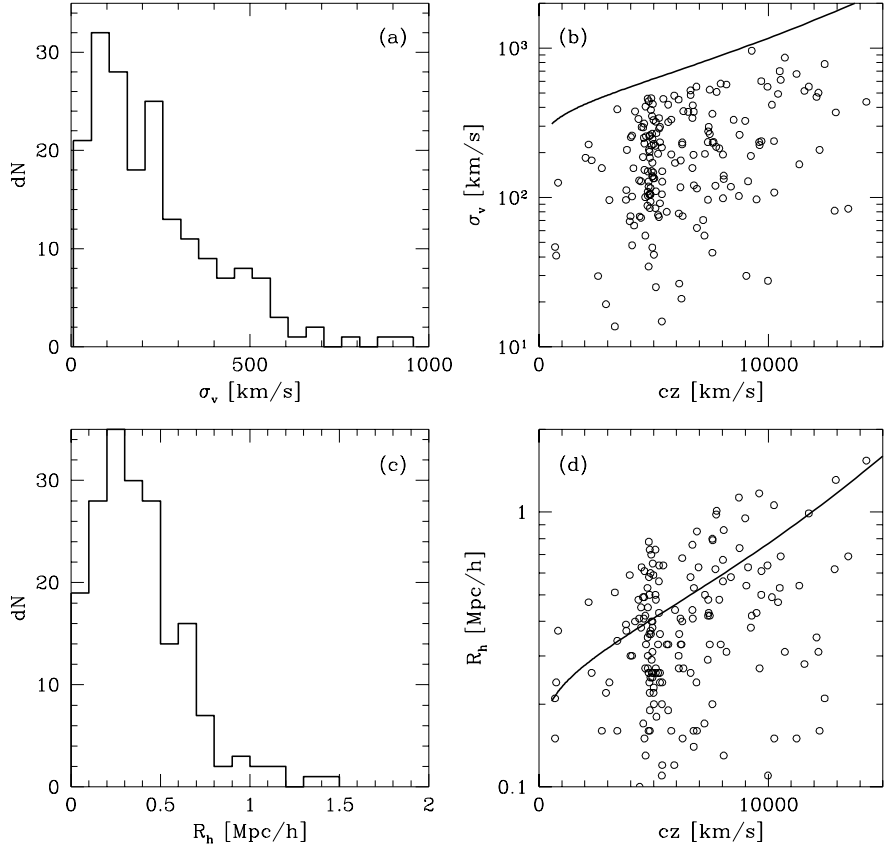}\end{minipage} 
\end{center} 
\caption{\label{fig7abcd}
As for Fig.~6, but here
(a,b) $X=\sigma_v$, the smooth curve is $V_L$;
(c,d) $X=R_h$, the smooth curve is $D_L$.
The curves are not fitted to the data.
They are directly obtained 
from the FOF links and the definition of the $X$'s (see text),
by replacing 
the harmonic radius $R_h$ and the line--0f--sight velocity dispersion $\sigma_v$ 
with the transverse spatial link $D_L$ and the radial velocity link $V_L$, respectively.}
\end{figure*}

\begin{figure*}
\begin{center} 
\epsfxsize=12cm
\begin{minipage}{\epsfxsize}\epsffile{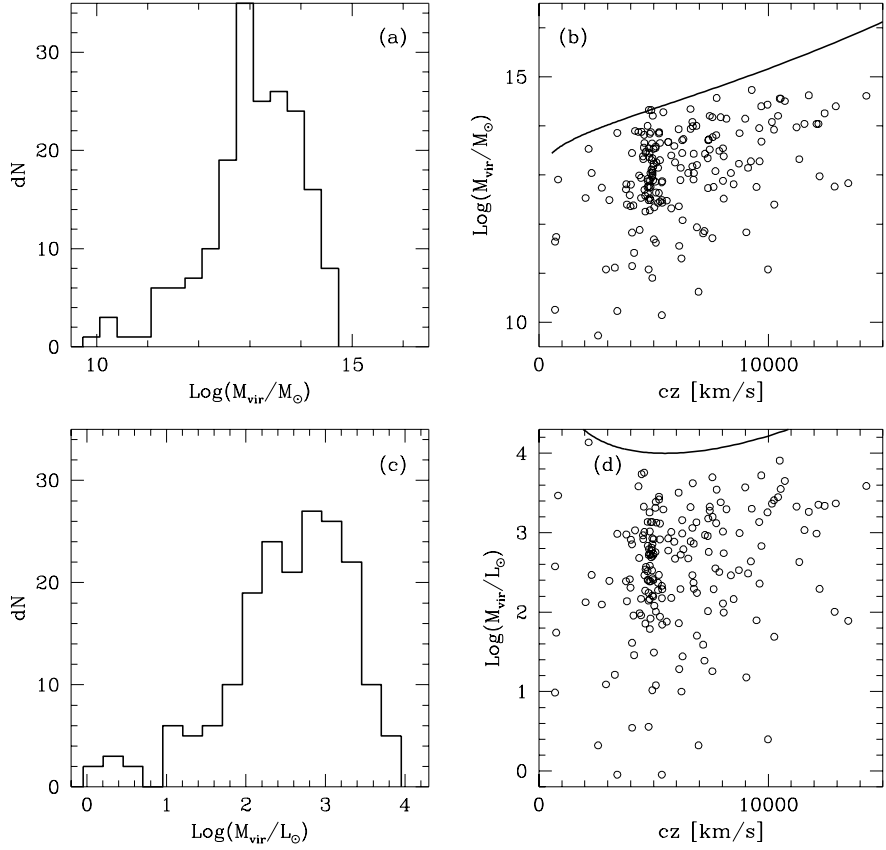}\end{minipage} 
\end{center} 
\caption{\label{fig8abcd}
As for Fig.~7, but here
(a,b) $X=\log_{10}(\Mvir/\M_\odot)$, 
the smooth curve is $6 \times G^{-1} \sigma_v^2 R_h$;
(c,d) $X=\log_{10}(\Mvir/L_G)$, 
the smooth curve is $2 \times G^{-1} \sigma_v^2 R_h /L_{\rm lim}$.
}
\end{figure*}

\begin{figure*}
\begin{center} 
\epsfxsize=12cm
\begin{minipage}{\epsfxsize}\epsffile[35 440 590 715]{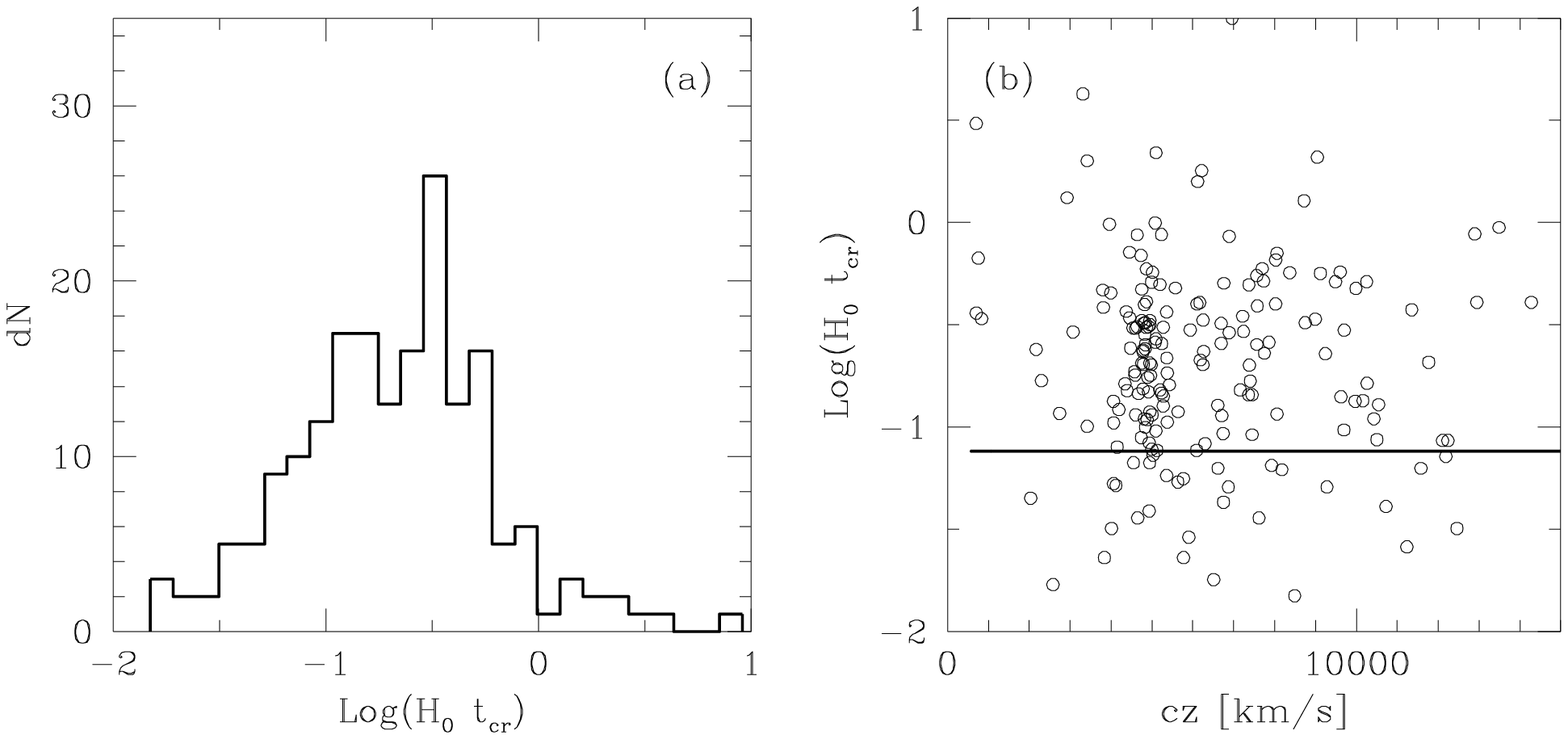}\end{minipage} 
\end{center} 
\caption{\label{fig9ab}
As for Fig.~6 and 7, but here
$X=\log_{10}(H_0 \tcr)$,
the smooth curve is $2/ \sqrt{3} \times D_L / V_L$.
}
\end{figure*}

\section{Loose groups in PPS2}
\label{sez:results}

\subsection{Group catalog}
\label{sez:catalog}

Here, we present the group catalog selected with $V_0=350 \ \kms$ and $D_0=0.231 \ \hMpc$,
($\alpha=-1.15$, $M_*=-19.30$, $\delta n/n = 173$ if $\phi_*=0.02 \ h^3 \ \Mpc^{-3}$),
which will be made available in electronic form at the CDS
(Centre de Donn\'ees astronomiques de Strasbourg, 
ftp://cdsarc.u--strasbg.fr, http://cdsweb.u-strasbg.fr/CDS.html).
The other catalogs, similarly selected with different parameter choices, 
may be obtained from the author upon request.

There are $N_G=188$ groups with $N_{mem} \geq 3$ members (105 with $N_{mem} \geq 5$)
corresponding to a total grouped fraction $f_{gr}=35 \%$.
There are 1406 singles ($47 \%$) and 283 binaries ($19 \%$).
Group members and group centroids are shown in 
Figures~\ref{fig3mem} and \ref{fig4grp}, respectively,
using the ``4+1--diagram'' described in Sect.~\ref{sez:data}.

Coordinates and internal properties of each group are given
in the electronic tables enclosed to the present paper.
As an example, 
in Table~\ref{tab_catgrp} 
we show the first and the last line of the group catalog.
For each group, we list a label,
the coordinates of the group centroid, and group internal properties.
All means are number--weighted,
$\langle \ \ \rangle$ denotes averages over pairs with $i \not =j$.  
The Table columns are:
(1) group identification number; 
(2) number of observed members $N_{mem}$;
(3) mean right ascension $\alpha_{1950}$; 
(4) mean declination $\delta_{1950}$; 
(5) mean redshift $cz$, MBR--corrected;
(6) r.m.s. velocity dispersion $\sigma_{v}$ (line--of--sight);
(7) 
$R_p = (cz /H_0) (4/\pi) 2 \sin (\langle \theta_{ij}\rangle/2)$,
mean pairwise member separation;
(8) 
$R_{\rm h}= (cz /H_0) (\pi/2) 2 \sin (\langle \theta_{ij}^{-1} \rangle^{-1}/2)$,
mean harmonic radius;
(9) $\log_{10}$ of the total blue luminosity 
$L_G = \sum L_i$ of all observed members,
extinction--corrected, assuming $M_{\odot}=+5.48$ (blue);
(10) $\log_{10}$ of the virial mass $\Mvir=6 G^{-1} \sigma_{v}^2 R_{\rm h}$;
(11) $\log_{10}$ of the virial--mass--to--observed--luminosity ratio $\Mvir/L_G$;
(12) crossing time $\tcr = 2 R_{\rm h} / \sqrt{3} \sigma_{v}$.
Definitions are the same as in RGH89, except $\tcr$ as in NW87,
whose numerical factor is a factor 4.30 larger than in Eq.(11) of RGH89.
Celestial coordinates  $\alpha$ and $\delta$
(in hours, minutes and fractions and 
degrees, minutes and fractions, respectively),
are given for the epoch 1950,
as in the original PPS database and in most group catalogs.
Velocities are in $\kms$, distances in $\hMpc$, 
masses in $h^{-1}\M_\odot$, luminosities in $h^{-2}L_\odot$.

For completeness, we provide the ratio $\Mvir/L_G$ as in other group catalogs,
but we regard its physical interpretation with caution, for the following reasons.
The physically interesting mass--to--light ratio
involves the {\emph true} group mass and its {\emph total} luminosity $L$. 
The former is usually estimated assuming the virial theorem to hold
for loose groups, which is probably not the case
(Aarseth \& Saslaw 1972; Giuricin et al. 1984, 1988; Heisler et al. 1985; 
Mamon 1993, 1996a; F95b; NKP97).
The {\emph total} luminosity $L$ of a group could be estimated 
from the {\emph observed} portion $L_G$,
the group richness $N_{mem}$, and the galaxy LF
(Gott \& Turner 1977; Bahcall 1979;
Mezzetti et al. 1985; NW87; Gourgoulhon et al. 1992; Moore et al. 1993).
Unfortunately, this would introduce further uncertainties and 
a further sample--to--sample dependence of group properties on $\phi(L)$
(other than that -- physical -- due to the different galaxy mixture,
and the one -- observational -- due to the FOF identification procedure).

\begin{table*}
\begin{center}
\caption{\label{tab_medetc} Global properties of loose groups in PPS2}
\begin{tabular}{l r r    r r r    r r r r}
\hline 
 & & 	& & &	& & & &	\\

Statistic
 		&$N_{mem}$	&$cz$
&$\sigma_v$ 	&$R_p$   	&$R_h$    
&$\log(L_G)$ 	
&$\log(\Mvir)$  	
&$\log({\Mvir \over L_G})$
&$\tcr$ \\
 		&	    	&$\kms$
&$\kms$     	&$\hMpc$ 	&$\hMpc$  
&
& 
&
&$H_0^{-1}$ \\ 

 & & 	& & &	& & & &	\\
\hline
 & & 	& & &	& & & &	\\
median 	&4.0	&5350	&194	&0.41	&0.34	&10.55	&13.16 	&2.67	&0.22\\
average	&5.5	&6250 	&232	&0.47	&0.39	&10.53	&13.06 	&2.54	&0.21\\
minimum	&3.0	&700	&7	&0.02	&0.00	&9.07	&9.73  	&-0.07	&0.15\\
maximum	&57.0	&14300 	&960	&2.26	&1.54	&11.63	&14.74 	&4.14	&4.26\\
$1^{st}$ 
quartile &3.0	&4800 	&102	&0.29	&0.21	&10.31	&12.58 	&2.16	&0.11\\
$3^{rd}$ 
quartile &6.0	&8050	&330	&0.61	&0.53	&10.76	&13.78 	&3.12	&0.40\\
($3^{rd}$q.$-1^{st}$q.)$/2$
	  &1.5	&1600 	&114	&0.16	&0.16	&0.23	&0.40	&0.48	&0.15 \\
stand.dev.&6.3	&2550 	&176	&0.31	&0.26	&0.38	&0.38 	&0.78	&0.26\\
stand.dev.
$/ \sqrt{N_G}$
	&0.5	&187 	&13	&0.02	&0.02	&0.03	&0.03	&0.06	&0.02 \\
 & & 	& & &	& & & &	\\
\hline
\end{tabular}
\end{center}
\end{table*}

\subsection{Group properties}
\label{sez:properties}

External properties of members and groups
(e.g., spatial position, clustering, grouped fraction...)
are easily visualized in Figs.~\ref{fig3mem} and \ref{fig4grp}.
They should be compared with the parent galaxy sample (Fig.~\ref{fig1gal}).
Our catalogs are built using all galaxies in PPS2. Previous works
(RGH89; F95; RPG97) cut their subsamples at  $cz \le 12000$-$15000 \ \kms$,
in order to exclude unreasonably elongated groups.
Tests with similarly cut subsamples in PPS show negligible differences.
In practice, the two procedures are equivalent,
as in the faraway regions there are too few galaxies to be grouped
(compare Fig.~\ref{fig1gal}, \ref{fig3mem}, and \ref{fig4grp}).
We further compare the radial distribution of galaxies, members, and groups
in Figure~\ref{fig5histo}.
The ratio among 
the observed number densities $\tilde n_G$ of FOF--identified groups
and $\tilde n_g$ of galaxies in PPS2
is rather constant and independent of redshift,
$\tilde n_G(r)/ \tilde n_g(r) \sim 1/15$ with our chosen FOF.
Consistent with this result, the total number ratio in CfA2 North is
$N_G/N_g \sim 6\%$ (RGH89; PGHR94; RPG97).

Internal properties 
($X=N_{mem}$, $L_G$, $\sigma_{v}$, $R_{\rm h}$, $\Mvir$, $\Mvir/L_G$, $\tcr$), 
are shown in Figures~\ref{fig6abcd},\ref{fig7abcd},\ref{fig8abcd}, \ref{fig9ab},
as distribution histograms $dN(X)$ and 
scatter plots $cz$-$X$ against redshift.
Table~\ref{tab_medetc} lists typical values and range of variability 
(average, median; minimum and maximum, $1^{st}$ and $3^{rd}$ quartile;
half interquartile range, r.m.s. deviation, r.m.s. deviation$/\sqrt{N_G}$)
of group properties.

\begin{table*}
\begin{center}
\caption{\label{tab_medLF}
Group properties (medians) for different 
combinations of LF, $D_0$, and ${\delta n \over n}$}
\begin{tabular}{l l l   r r r r r r r r  r}
\hline
 & &	  & & & & & & & & 	&\\
$\phi(M)$ &$D_0$ &${\delta n \over n}$
&$N_G$	&$cz$ &$\sigma_v$ &$R_p$ 
&$R_h$ &$L_G$ &$\Mvir$ &$\Mvir/L_G$ &$\tcr$\\
          &$\hMpc$     &	
&       &$\kms$ &$\kms$ &$\hMpc$ &$\hMpc$ 
&$h^{-2} L_\odot$ &$h^{-1}\M_\odot$ &$M_\odot L_\odot^{-1}$ &$H_0^{-1}$\\ 
 & & 	  & & & & & & & &	&\\
\hline
 & & 	  & & & & & & & &	&\\
TB96 &0.231 &173 
&188 &5350 &194 &0.41 &0.34 &3.52$\times 10^{10}$ &1.44$\times 10^{13}$ &470 &0.22\\
RPG97 &0.231 &80  
&201 &5650 &193 &0.48 &0.39 &3.78$\times 10^{10}$ &1.97$\times 10^{13}$ &540 &0.23\\
TB96  &0.270 &108 
&201 &5500 &193 &0.51 &0.41 &3.78$\times 10^{10}$ &1.90$\times 10^{13}$ &520 &0.26\\
RGH89 &0.270 &80  
&211 &6700 &193 &0.57 &0.43 &4.18$\times 10^{10}$ &1.85$\times 10^{13}$ &530 &0.26\\
TB96  &0.300 &79  
&210 &5630 &187 &0.59 &0.47 &3.99$\times 10^{10}$ &1.81$\times 10^{13}$ &520 &0.29\\
 & &	  & & & & & & & & 	&\\
\hline
\end{tabular}
\end{center}
\end{table*}

\begin{table*}
\begin{center}
\caption{\label{tab_medcz}
Group properties (medians) for different redshift corrections}
\begin{tabular}{l r r    r r r    r r r}
\hline
 & & 	& & &	& & & 	\\
Rest Frame
&$N_G$	&$cz$ &$\sigma_v$ &$R_p$ &$R_h$ &$L_G$ &$\Mvir$ &$\tcr$ \\
&       &$\kms$ &$\kms$ &$\hMpc$ &$\hMpc$ &$h^{-2} L_\odot$ &$h^{-1}\M_\odot$ &$H_0^{-1}$\\ 
 & & 	& & &	& & & 	\\
\hline
 & & 	& & &	& & & 	\\
$D_0=0.231 \ \hMpc$ 
 & &	& & &	& & & 	\\
 & &	& & &	& & & 	\\
MBR &188 &5350	&194 &0.41 &0.34 &3.51$\times 10^{10}$ &1.44$\times 10^{13}$ &0.22\\
SUN &187 &5450	&198 &0.42 &0.36 &3.56$\times 10^{10}$ &1.54$\times 10^{13}$ &0.22\\
LGC &186 &5550	&193 &0.43 &0.36 &3.80$\times 10^{10}$ &1.49$\times 10^{13}$ &0.22\\
 & &	& & &	& & & 	\\
$D_0=0.270 \ \hMpc$ 
 & &	& & &	& & & 	\\
 & &	& & &	& & & 	\\
MBR &201 &5500	&193 &0.51 &0.41 &3.78$\times 10^{10}$ &1.90$\times 10^{13}$ &0.26\\
SUN &205 &5650	&193 &0.51 &0.42 &3.96$\times 10^{10}$ &1.87$\times 10^{13}$ &0.27\\
LGC &204 &5750	&192 &0.52 &0.43 &4.07$\times 10^{10}$ &1.64$\times 10^{13}$ &0.27\\
 & & 	& & &	& & & 	\\
\hline
\end{tabular}
\end{center}
\end{table*}

\subsection{Variability of group properties}
\label{sez:variability}

Group catalogs selected from different galaxy samples
might be significantly inhomogeneous with each other even
when the parent galaxy samples are homogeneously selected. 
The primary source of discrepancy would be, of course,
an inconsistent matching of FOF parameters among different catalogs,
i.e. 
(i)~a different link normalization for a given sample depth, or
(ii)~a different sample depth for a given link normalization
(see Sect.~\ref{sez:normalization}).
Further, subtler sources of discrepancies could be:
(iii)~a different galaxy LF, which depends on physical differences among samples,
but also plays an active role in the FOF algorithm itself;
(iv)~large scale flows of peculiar motions;
(v)~sample--to--sample variations, e.g. due to local LSS features within the samples.

In Table~\ref{tab_medLF}, we test for variations of internal properties
due to different assumptions about the galaxy LF.
We consider several group catalogs 
with several combinations of LF, $D_0$, and $\delta n/n$,
but the same $V_0=350 \ \kms$.
(We normalize the LF of TB96 with $\phi_*=0.02 \ h^3 \ {\rm Mpc}^{-3}$.)
The main differences are connected with the different $D_0$.
For given $V_0$ and $D_0$, the residual net effect of the galaxy LF 
on group internal properties is generally rather small, 
$\delta X/X \simlt 5$-$10 \%$.
Similar results hold also for group positions and clustering properties (TB96; TBIB97).

In Table~\ref{tab_medcz}, we test the effect of different $cz$ corrections 
for the motion of the Sun (none, MBR, local group centroid) 
described in Sect.~\ref{sez:data}.
As expected, we find negligible differences in group properties and membership
in the three cases (except, of course, 
an overall modulation of $cz$ of group members and centroids).
In particular, no--correction or MBR--correction
yield almost indistinguishable results,
as the direction of $v_{\odot MBR}$ ($\alpha=11.2^h$; $\delta=-7^o$)
is almost orthogonal to (the bulk of) PPS.

Table~\ref{tab_PPSCfA} is a preliminary comparison of our group catalog with 
similarly selected groups in previous studies.
We note explicitly that all these samples have the same depth, given by $\mlim=15.5$.
A more thorough investigation (Trasarti--Battistoni 1997)
is beyond the scope of this paper.
We list global properties for our groups in PPS2 and for
groups in the CfA2 survey as given in RGH89, F95b, PGHR94, and RPG97.
Their results and ours are in good agreement.
Comparing our Table~\ref{tab_medetc} with Table~6 of RGH89 and
Table~1 of PGHR94 shows that
also the ranges of variability of group properties
are in very good agreement in the three cases.

\begin{table*}
\begin{center}
\caption{\label{tab_PPSCfA} 
Group properties in PPS2 and in CfA2 (medians, and global values)}
\begin{tabular}{l l l l l l l}
\hline
         &           &        &           &                 &      &        \\
Galaxy sample   &$\sigma_v$ &$R_h$   &$t_{cr}$   &$\M_{vir}$ &$N_G/\omega$ &$f_{gr}$\\
         &$\kms$     &$\hMpc$ &$H_0^{-1}$ &$h^{-1}\M_\odot$ &${\rm sr}^{-1}$  & \\ 
         &           &        &           &                 &      &        \\
\hline
         &           &        &           &       &                    &    \\
$D_0=0.231 \ \hMpc$ 
         &           &        &           &       &                    &    \\
         &           &        &           &       &                    &    \\
PPS2	      &194   &0.34    &0.22       &1.44$\times 10^{13}$ &2.5$\times 10^{2}$  &0.35\\
CfA2 North    &192   &0.40    &0.21       &1.23$\times 10^{13}$ &3.4$\times 10^{2}$  &0.40\\
         &           &        &           &       &                    &    \\
$D_0=0.270 \ \hMpc$ 
         &           &        &           &       &                    &    \\
         &           &        &           &       &                    &    \\
PPS2	      &193   &0.41    &0.26       &1.90$\times 10^{13}$ &2.6$\times 10^{2}$  &0.41\\
CfA2 Slice(s) &215   &0.41    &0.22       &2.57$\times 10^{13}$ &3.0$\times 10^{2}$  &0.44\\
         &           &        &           &       &                    &    \\
\hline
\end{tabular}
\end{center}
\end{table*}

This seems to contradict the results of RGH89.
They found a significant difference among groups in different samples,
namely $\sigma_{v, {\rm med}}=131 \ \kms$ in the CfA1 survey 
and $\sigma_{v, {\rm med}}=192 \ \kms$ in the CfA2 Slice. 
From their Figure~9, one sees that groups in the CfA1 survey and the CfA2 Slice 
are located 
preferentially around $cz_1 \approx 1000 \ \kms$ and $cz_2 \approx 8000 \kms$,
respectively.
RGH89 correctly notice that this different location of the LSS within the samples
induces a significant physical difference among the groups in the two samples, 
those in the shallower sample being typically nearer to us and brighter. 
Then, they argue that these sample--to--sample
variations might also be responsible for the discrepancy in $\sigma_{v, {\rm med}}$.
However, RGH89 do not give an explanation why $\sigma_{v, {\rm med}}$ should be
higher in one sample than in the other, and they reject the possibility
that the discrepancy be induced by the FOF grouping algorithm.
(See also the discussion in Maia et al. 1989).
In fact, the difference between the groups in CfA1 and CfA2 Slices
could be due to a {\emph combination} of 
the different LSS features present in the two galaxy samples and of
the different radial scaling of the FOF links adopted for the two samples.
We recall that the links increase with $r$ and decrease with $\mlim$ (Sect.~\ref{sez:scaling}).
Therefore, normalizing the links with the same $D_0$ and $V_0$ for both samples,
but scaling them proportionally to $\left[ \bar n(r; \mlim) \right]^{-1/3}$ 
(different for the two samples!) as in RGH89,
at any given $r$ the links $D_L$ and $V_L$ 
will still be always more generous in the shallower sample than in the deeper one.
But, if group properties are so directly related to $D_L$ and $V_L$ as suggested
by Sect.~\ref{sez:normalization}, then the relative location of LSS features
within the sample boundaries must {\emph also} be taken into account.
E.g., if the LSS features lie at comparable distance in both samples,
one would expect a higher $\sigma_{v, {\rm med}}$ in the shallower sample (more generous links).
But, if the LSS features are very differently distributed in the two samples,
they could be differently ``weighted'' by the adopted links.
So, if LSS features lie at a much greater distance (more generous links)
in the deeper than in the shallower sample -- as in the case for CfA1 and CfA2 Slices --
scaling up the links with $r$ might even overtake the effect of $\mlim$ in $\bar n(r;\mlim)$.
One would then expect a higher $\sigma_{v, {\rm med}}$ in the deeper sample -- 
once more, as in the case for CfA1 and CfA2 Slices.
Interestingly, one finds a link ratio $V_L(cz_1;{\rm CfA1})/V_L(cz_2;{\rm CfA2}) \approx 2.5$,
to be compared with 
$\sigma_{v, {\rm med}}({\rm CfA1})/\sigma_{v, {\rm med}}({\rm CfA2})\approx 1.5$,
taking into account that groups are not only located precisely at $cz_1$ and $cz_2$.
In a sense, by modulating the properties of the groups according to distance,
the FOF links may either amplify or deamplify the sample--to--sample variations
according to how the galaxy LSS is arranged within the samples. 
On the other hand, such effects will be reduced 
when samples of the same $\mlim$ are compared using the same $D_L(r)$ and $V_L(r)$, as we do here.
In this case, all discrepancies would be purely due to the intrinsic sample--to--sample variations,
i.e. a different amount and/or location of LSS within the survey limits,
but not further modulated by a different radial scaling of the links.
Considering medians or averages over the whole group distributions
would further reduce the sample--to--sample discrepancies.

Related to the previous point, note the smooth curves in the $cz$-$X$ scatter plots 
(Fig.~\ref{fig6abcd}--\ref{fig9ab}). 
They were {\emph not} obtained by fitting the observed distribution on the diagrams.
They were obtained simply 
by replacing $\sigma_{v}$ and $R_{\rm h}$ by $D_L(cz)$ and $V_L(cz)$, respectively,
in all formulae defining internal properties.
However, there is often a clear similarity of redshift dependence 
between the smooth, FOF--induced curves 
and the (upper envelopes, or median values of) group internal properties.
Note also how the peaks in the $dN(X)$ histograms often correspond to 
denser region in the $cz$--$X$ plane (projected onto the $X$ axis), 
in turn related with dense concentrations in redshift space 
(e.g., the peak at $cz\sim 5000 \ \kms$ in 
Fig.~\ref{fig1gal}, \ref{fig3mem} and \ref{fig4grp}).

In summary, in PPS2 as well as in other samples (Maia et al. 1989; RGH89),
there seem to be a complex interplay among LSS features, sample depth,
FOF algorithms, and group properties.
Disentangling these effects is a subtle matter,
out of the scope of the present paper,
and it is left for a future work (Trasarti--Battistoni 1998, in preparation).

\section{Conclusion}
\label{sez:conclusion}

This paper had one key aim:
to build several large group samples ($N_G\approx 200$ groups) 
in the Southern Galactic Hemisphere
from the PPS galaxy survey, never previously analyzed in this way.
Such galaxy sample is considerably 
larger and/or deeper and/or wider than those used in most similar previous studies,
so that our group catalog is one of the largest presently available.

Particular care was used in order to define group catalogs
as homogeneous as possible to those previously published
-- in particular, the large group catalogs 
based on the CfA2 galaxy survey in the Northern Galactic Hemisphere (RGH89; RPG97).
Such samples have the same depth as our sample PPS2, and comparable angular width, 
but different galaxy LF. 

Group catalogs are customarily labelled by the redshift link $V_0$
and the effective density contrast threshold $\delta n/n$ used to select the groups
(or, equivalently, the mean inter particle separation $\bar n^{-1/3}$).
However, to specify spatial separations, the parameter actually used by the FOF algorithm 
is not $\delta n/n$, but rather the spatial link $D_0$ itself. 
The relations among these two parameters depends on the adopted galaxy LF and sample depth, 
so it differs from sample to sample.
This leads to some ambiguity, and to several possibilities, which we discuss, 
about how to match our grouping algorithm to those used for the other samples.
On one hand, and consistently with Maia et al. (1989) and RPG97,
we find strong, approximately linear correlations 
(i) between the redshift link $V_0$ and the (median values of) 
the velocity dispersion $\sigma_{v}$,  
and 
(ii) between the spatial link $D_0$ and the (median values of) 
the mean harmonic radius $R_{\rm h}$ and mean pairwise member separation $R_p$.
Even for individual groups, 
the redshift dependence of $R_{\rm h}$ and $\sigma_{v}$ 
seems to be closely related to $D_L$ and $V_L$, respectively.
On the other hand, and consistently with RPG97,
group velocity dispersions (spatial sizes) are rather insensitive
to the spatial link $D_L$ (velocity link $V_L$).
All this suggests to regard $D_0$ and $V_0$ as the basic FOF parameters,
and interpret $\delta n/n$ only as an estimate of the density contrast threshold.

We adopt the normalizations $D_0=0.231 \ \hMpc$ and $V_0=350 \ \kms$, as in RPG97.
The galaxy LF for PPS2 has Schechter parameter (STY fit) 
$\alpha=-1.15 \pm 0.15$ and $M_*=-19.3 \pm 0.1$, in good agreement with similar estimates.
The STY technique does not allow to estimate the LF normalizations $\phi_*$.
We then adopt the value $\phi_* =0.02 \ \pm 0.1 \ h^3 \ \Mpc^{-3}$
as determined by MHG94 for the CfA2 South sample, which is very similar to PPS2.
The adopted normalizations and LF yield then $\delta n/n=173$.
We test for the effect of galaxy LF on group properties.
The main effect is connected to the relation between $D_0$, $\delta n/n$, and $\phi_*$,
and the uncertainty on the latter.
By replacing the $\delta n/n$ parametrization with the $D_0$ parametrization,
this problem is avoided.
In fact, for given $V_0$ and $D_0$, the residual net effect on group properties 
due to $\alpha$ and $M_*$ is generally small:
$\delta X/X \simlt 5$-$10 \%$ for all considered internal properties $X$,
and similarly for group positions (TB96) and group clustering (TBIB97).
We also test for the effect of different redshift corrections. 
Again, the effect is small, as expected for magnitude--limited samples.

Our main conclusions are as follows:

1. The spatial distribution of FOF--identified loose groups in PPS2
largely reproduce the LSS features in the parent galaxy catalog.
Thus, galaxy loose groups can be usefully used as tracer of LSS.
Analysis of group clustering in PPS2 has been presented elsewhere (TBIB96). 

2. Properties of FOF--identified loose groups selected from 
directly comparable (in depth, selection criteria, sky coverage, etc.)
parent samples are generally in good agreement,
provided group are selected in a similar way.

3. However, there seems to be a complex interplay among 
the LSS features in the galaxy sample, the sample depth, 
the FOF grouping procedure, and the group properties.
A more detailed assessment of this and the previous point 
will be presented elsewhere (Trasarti--Battistoni 1998, in preparation).

The large extent of the group catalog presented here is due to the
depth, sky coverage, and high sampling density of the parent galaxy sample PPS.
The deep, high--density, and wide--angle surveys 
CfA2 and SSRS2 have been completed already some years ago,
and they should be made available in the future (Ramella, private communication).
These samples are directly comparable to PPS2, 
and we hope that they will be suitably combined with it for future analysis.
The group catalog presented here was built with this purpose in mind.

Much larger samples will be required for further, substantial improvement.
In fact, the deeper surveys nowadays available
are usually not well--suited to group analysis.
Infrared--selected surveys (e.g., Fisher et al. 1995)
contain preferentially late--type galaxies,
thus biased against high density regions,
and their infrared LF yields a SF rapidly decreasing with $cz$, 
in this way exhacerbating the scaling problem.
Very deep surveys, 
sparse samples (e.g., Loveday et al. 1992) or
narrow angle surveys (e.g., Vettolani et al. 1993), 
add extra difficulties to this kind of study, 
as group identification require a sampling ratio as high as possible,
and it is more difficult to identify groups near the survey edges.
Future surveys such as
2dF (Colles \& Boyle 1998), 6dF (see Mamon 1996b),
and SDSS (Gunn \& Weinberg 1995), 
will provide homogeneous galaxy samples
(250\,000 in 2 slices by 1999; 
90\,000, near--IR selected, over the southern sky by 2002;
1\,000\,000 over half the northern hemisphere by 2004, respectively)
that should provide considerably larger homogeneous catalogs of loose groups.

\paragraph{Acknowledgements} 

I am glad to thank 
R.~Giovanelli \& M.P.~Haynes, who kindly supplied a computer version
of their data and carefully commented an early version of this work.
The LUMFUN package (Bardelli et al. 1990) was kindly provided by E.~Zucca.
I am grateful to the referee G.~Mamon, for his continous suggestions
who considerely helped me to improve the presentation of this work.
Very helpful suggestion came also from M.~Ramella, to whom I am also grateful.
Special thanks are due to R.~Nolthenius, 
for his precise and precious criticisms on an early version of this work.
I also received useful suggestions from
B.~Bertotti, S.A.~Bonometto, M.J.~Geller, J.R.~Primack, 
and the ``$P + GM^2/R$'' group at SISSA. 
Finally, my personal thanks go to
A. Diaferio, S. Ettori, A. Gardini, S. Ghizzardi, 
G. Giudice, F. Governato, G. Invernizzi, and R. Mignani.
This research was supported by the Italian MURST.



\end{document}